# RESEARCH NOTE ON A PARABOLIC HEAT-BALANCE INTEGRAL METHOD WITH UNSPECIFIED EXPONENT:
## An Entropy Generation Approach in Optimal Profile Determination

by

*Jordan* **HRISTOV**

*The heat-balance integral method of Goodman is studied with two simple 1-D heat conduction problems with prescribed temperature and flux boundary conditions. These classical problems with well known exact solutions enable to demonstrate the heat-balance integral method performance by a parabolic profile and the entropy generation minimization concept in definition of the appropriate profile exponent. The basic assumption generating the additional constraints needed to perform the solution is based on the requirement to minimize the difference in the local thermal entropy generation rates calculated by the approximate and the exact profile, respectively. This concept is easily applicable since the general concept has simple implementation of the condition requiring the thermal entropy generations calculated through both profiles to be the same at the boundary. The entropy minimization generation approach automatically generates the additional requirement which is deficient in the set of conditions defined by the heat-balance integral method concept.*

## Introduction

Heat-balance integral method (HBIM) of Goodman [1] is an effective method for solving heat diffusion problems with strong non-linearity either in the energy equation or at the boundaries. The basic idea lies on a physically-based formulation of a thermal layer (this avoids the inadequacy of the Fourier equation) and a prescribed temperature profile. These essential ideas allow transformation of strong non-linear 1-D heat conduction problems into ordinary differential equation with respect to the thermal layer temperature evolution [1, 2]. The common approach is to use polynomial temperature approximations with respect to the space co-ordinate with up to 4 boundary conditions allowing defining the profile coefficients as functions of the thermal layer depth – see eqs. (3a, b) too – namely:

$$T(0,t) = T_s \quad \text{or} \quad \lambda \frac{\partial T}{\partial x} = \dot{q}_s, \text{ at } x = 0 \text{ (the face of the heated body)} \quad (1a, b)$$

$$T(\delta,t) = T_\infty, \; \lambda \left.\frac{\partial T}{\partial x}\right|_{x=\delta} = 0, \text{ or } \left.\frac{\partial^2 T}{\partial x^2}\right|_{x=\delta} = 0 \text{ at } x = \delta(t) \text{ (the front of the thermal layer)} \quad (1c, d)$$

with

$$\delta(t = 0) = 0 \quad (2)$$

where $\delta(t)$ is the depth of the thermal penetration layer; the crux of the Goodman's method [1, 2] based on physics of the heat conduction.

The conditions (1a) and (1b) are classical for the prescribed temperature and prescribed flux problem, respectively. The fourth condition (1d) is know as "smoothing condition" and works well when a polynomial approximation of $4^{th}$ order is used. Generally, the accuracy of the HBIM depends on the adequate choice of the approximating functions and the literature provided many examples of successful solutions [3-10]. The present work focus the attention on HBIM solution of heat-conduction problems (isothermal condition and prescribed flux problems):

$$\frac{\partial T(x,t)}{\partial t} = \alpha \frac{\partial^2 T(x,t)}{\partial x^2}, \quad 0 \le x \le \delta(t) \quad (3a, b)$$

where $T(x, t)$ is preliminarily defined parabolic profile with unspecified exponent, namely:

$$T(x, t) = a + b(1 + cx)^n \quad (3b)$$

The profile (3b) is very often used in the form (with $T(0, t) = T_s$):

$$\Theta = \frac{T(x,t) - T_\infty}{T_s - T_\infty} = \left(1 - \frac{x}{\delta}\right)^n \quad (3c)$$

Examples and analysis are provided by [11] where a clear algorithm to define the exponent $n$ through additional constraints based on exact solutions was developed. This approach allowed to obtain in a clear manner previously obtained solutions of specific cases such as those solved by Braga *et al.* [12]. Further, applying the Langford criterion [13] for error estimation yielded equations consistent with those from the additional physical constraints giving directly the optimal value of the exponent $n$. Now, we address a physically based approach named entropy generation minimization (EGM) in both the error estimation and the optimal exponent determination.

Some preliminary thoughts prior to the problem statement give the basic ideas coming from different areas of physics and modern thermodynamics. First of all, the EGM approach conceived by Prigogine [14] and Bejan [15, 16] is extensively applied to various problems commonly known as thermodynamic optimization [17] and entropy generation analysis [18-21]. As to the approximate solution of heat-conduction problems we credit to Esfahani [22] who has performed a direct comparison of the entropy generation provided by exact and approximated solutions to 2-D heat conduction problems with internal heat generation and his steps, therefore, will be briefly outlined next.

The one-way destruction of the useful work is directly propositional to the rate of entropy generation [15]:

$$W_{lost} = T_\infty S_{gen} \quad (4)$$

where $T_\infty$ is the absolute temperature of the ambient reservoir ($T_\infty$ = const.). Assuming a finite size control volume in the 1-D problem and applying the second law of thermodynamics, the mix entropy generation per unit time and per unit volume ($S_{gen}$), *i. e.* the local entropy generation after Bejan [15] is:

$$S_{gen} = \frac{\lambda}{T^2}\left(\frac{\partial T}{\partial x}\right)^2 + \frac{\mu}{T}\left(\frac{\partial V}{\partial y}\right)^2 + \frac{g}{T} \quad (5)$$

where $g$ is the volumetric energy source [W/m$^3$] and $V$ is the fluid velocity if a simple fluid flow through a pipe is considered as example. Since we address only heat-conduction problems only the heat energy terms of eq. (5) is at issue that yields simply the so-called local *thermal entropy generation* (TEG) [15, 22].

$$S_{gen} = \frac{\lambda}{T^2}\left(\frac{\partial T}{\partial x}\right)^2 + \frac{g}{T} \tag{6}$$

The TEG depends on the temperature profile and the function describing it which mainly is affected by the method of solution [22]. Henceforth we credit to Esfahani [22] who defined a dimensionless entry generation function which can be expresses for 2-D heat conduction ($h_0$ – plate length and $l_0$ – plate width) problem as:

$$\bar{S}_{gen} = \frac{S_{gen}}{\frac{\lambda}{h_0^2}} = \left(\frac{h_0}{l_0}\right)^2 \left(\frac{\partial \Theta}{\partial\left(\frac{x}{l_0}\right)}\right)^2 + \left(\frac{\partial \Theta}{\partial\left(\frac{x}{h_0}\right)}\right)^2 + \left(\Theta + \frac{T_\infty}{\frac{gh_0^2}{\lambda}}\right)^{-1} + \left(\Theta + \frac{T_\infty}{\frac{gh_0^2}{\lambda}}\right)^{-1} \tag{7}$$

The average normalized TEG is defined as:

$$\bar{S}_{gen(average)} = \frac{\int_A \bar{S}_{gen} dA}{\int_A dA} \tag{8}$$

The average error is defined as [22]:

$$\text{Error} = \frac{\int_A (\Theta_{exact} - \Theta_{approximate}) dA}{\int_A dA} \tag{9}$$

where d$A$ is a dummy (variable) area of integration.

The present article implements the idea of EGM as a measure of approximation error to solutions through the HBIM with the specific parabolic profile at issue. More exactly, the idea is not to calculate the error of approximation but to use the assumption that irrespective of the temperature profile (exact or approximate) used to calculate the local entropy generation higher accuracy of approximation should be assured if their difference goes to a minimum or at least to zero. The idea is that such a minimization could be performed through comparison of the global entropies or their local (nodal) values at specific points of the thermal penetration layer. This minimization procedure generates new conditions (constraints) affecting the approximate profile but not available in the set of original ones defined by HBIM. This idea is explained and outlined next.

**Problem statement and basic idea of TEG approach to HBIM**

Using TEG definition – see eq. (6) – the local entropy generation in absence of volumetric heat generation [23] is:

$$\sigma(T, T_x) = \frac{dS}{d\tau} = \frac{\lambda}{T^2}(\text{grad}T)^2 \tag{10}$$

where $T_s = \partial T/\partial x$.

The general problem can be formulated in the following way: Find such a function $T(x, t)$ which satisfying required boundary conditions minimizes simultaneously the integral:

$$\sigma_t \quad \int_\Omega \sigma(T, T_s) d\Omega \tag{11}$$

over the whole domain $\Omega$.

The problems has been discussed extensively in [23] for steady-state 1-D heat conduction problems in view of adequacy of heat transfer equation derived through entropy minimization approach and the Fourier equation. Now, this is not the case but some results of [23] will be used to outline the problem at issue, namely:

(1) the entropy generation rate attains its minimum at steady-state irrespective of the way through which the function $T(x, t) \quad T(x) = T_\infty$ (see fig. 1a, b) is developed. Hence, at the approximate solution might provide a minimum of the entropy generation rate calculated through the approximate profile since at $x \quad \delta$ we have $\text{grad} T = 0$,

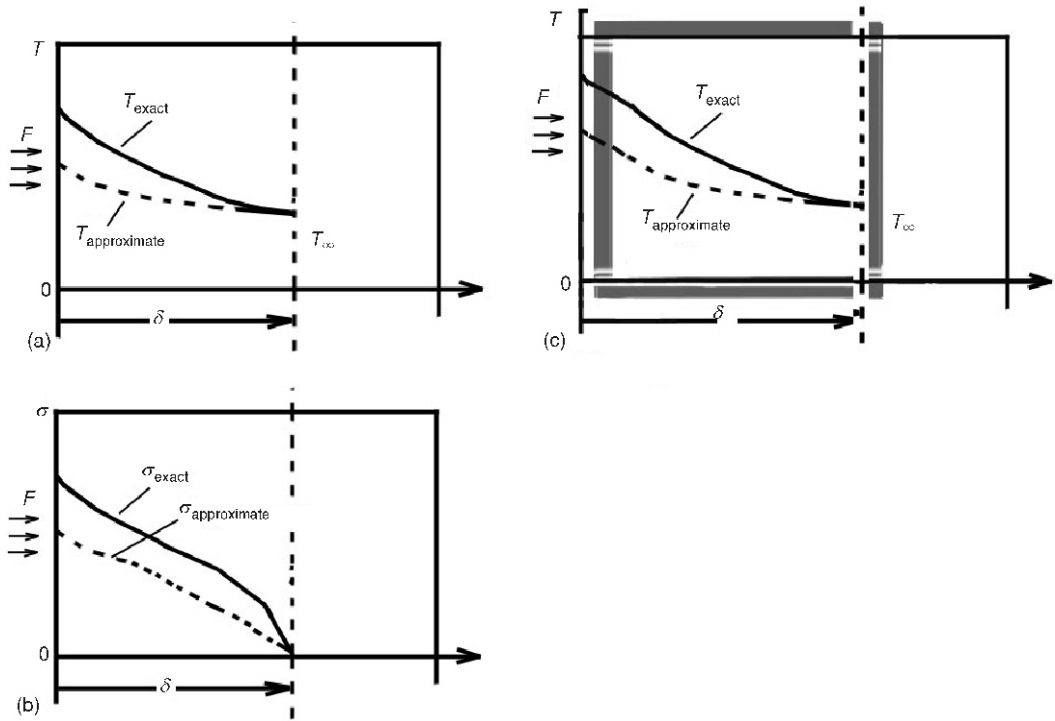

**Figure 1. Schematic presentation of the idea to use the entropy generation rate in determination of the optimal approximate temperature profile by HBIM**
*(a) Temperature profile across the thermal penetration layer; (b) Control volume (grey borders) based on the thermal penetration layer with a reference temperature at its walls ($T_s$ or $T_\infty$); (c) Local entropy generation profile through the thermal penetration layer expressed by the exact and the approximate solution. The entropy is positive and that expressed through the exact solution exhibits the local TEG minimum used a standard value*

(2) the exact solution also attains its minimum at steady-state ($t \quad \infty$) but we might suggest a basic principle that the TEG rate calculated through the exact solution is the minimal even though the heat transfer is time-dependent. This is implies that the TEG calculated through the exact solution is the standard and the approximated solution should approach it with a minimal error through adequate choice of the approximate profile,

(3) the above physically-based statements mean that, and

(4)
$$\Delta\sigma = \sigma(T, T_x)_a - \sigma(T, T_x)_e = 0 \quad (12a)$$

which implies a zero error of approximation in accordance with eq. (9) or at least:

$$\Delta\sigma = \sigma(T,T_x)_a - \sigma(T,T_x)_e = \text{minimum} \quad (12b)$$

*i. e.* a minimal error of approximation.

This approach will be used for creation of additionally physically-based constraints leading to optimal parabolic profile of HBIM which practically means a way to define the appropriate exponent *n*.

**Test of the concept through simple examples**

Several simple 1-D heat conduction problems will be used to exemplify the entropy generation approach in creation of additionally physically-based constraint leading the appropriate exponent.

Some of them were solved in [11] and the results obtained in this work will used directly in the development of the present analysis.

***Example 1.*** Prescribed temperature problem (PT) at $x = 0$

The HBIM solution of eq. (3a) with $T = T_s$ at $x = 0$ and $T(x, 0) = T_\infty$, provides [11]:

$$T(x,t) = T_\infty + (T_s - T_\infty)\left[1 - \frac{x}{\sqrt{\alpha t}\sqrt{2n(n+1)}}\right] \quad (13a)$$

with

$$\delta = \sqrt{\alpha t}\sqrt{2n(n+1)} \quad (13b)$$

Whilst *the exact solution* is [11, 24]:

$$T(x,t) = T_s + (T_s - T_\infty)\,\text{erf}\left(\frac{x}{2\sqrt{\alpha t}}\right) \quad (14)$$

The local entropy generation rate calculated through the approximate solution is:

$$\sigma(T,T_x)_a = \lambda \frac{(T_s - T_\infty)^2 \left(\frac{n}{\delta}\right)^2 \left(1 - \frac{x}{\delta}\right)^{2(n-1)}}{\left[T_\infty + (T_s - T_\infty)\left(1 - \frac{x}{\delta}\right)^n\right]^2} \quad (15a)$$

Whilst that calculated through the exact solution is:

$$\sigma(T,T_x)_e = \frac{(T_s - T_\infty)^2 \left(\frac{1}{\sqrt{\pi\alpha t}}\right)^2 \exp\left(-\frac{x^2}{2\alpha t}\right)}{\left[T_\infty + (T_s - T_\infty)\,\text{erf}\left(\frac{x}{2\sqrt{\alpha t}}\right)\right]^2} \quad (15b)$$

Then, the function $\sigma(T, T_x)$ has to be minimized through an adequate choice of the exponent $n$. The calculation $\Delta\sigma(T, T_x)$ of expressed through (12a) and the profiles (15a, b) and it minimization with respect to $n$ is complex and practically not necessary since two basic physical assumptions simplifying the solution should be formulated, namely:

(1) at $x = 0$ the local entropy generation rate, irrespective of the temperature profile used for its calculation, has a maximum – fig. 1a,
(2) at $x = \delta$ the local entropy generation rate, irrespective of the temperature profile used for its calculation, has a minimum. With the approximate profile, for instance, $\sigma(T, T_s)_a = 0$ (see 15b and fig. 1b) since the HBIM defines $T_s = 0$ at $x = \delta$, and
(3) therefore, if we may find a solution to minimize the difference between the differences between the maxima in TEG calculated by both profiles this would assure minimal differences over the entire thermal layer.

In accordance with the simplifying assumptions 1 and 3 we may assume the case when the approximate solution matches the exact one and the following condition is obeyed locally at $x = 0$:

$$\Delta\sigma = \Delta\sigma_{(x=0)} = \sigma(T,T_x)_a - \sigma(T,T_x)_e = 0, \quad x = 0 \tag{16}$$

This provides simple expressions of the function describing the TEGs, namely:

$$\sigma_{x=0}(T,T_s)_a = \left(\frac{n}{\delta}\right)^2 \left(\frac{T_s - T_\infty}{T_s}\right)^2 \tag{17a}$$

and

$$\sigma_{x=0}(T,T_x)_e = \left(\frac{1}{\sqrt{\pi}\sqrt{\alpha t}}\right)^2 \left(\frac{T_s - T_\infty}{T_s}\right)^2 \tag{17b}$$

The condition $\Delta\sigma_{(x=0)} = 0$, see eq. (16), and (17a), and (17b), yields:

$$\left(\frac{n}{\delta}\right)^2 = \left(\frac{1}{\sqrt{\pi}\sqrt{\alpha t}}\right)^2 \tag{18}$$

This gives the relationship:

$$\frac{\delta}{\sqrt{\alpha t}} = n\sqrt{\pi} \tag{19}$$

Recall, the relationship (19) was provided by the condition $q_a(x=0) = q_e(x=0)$ used in [11] with the same problem. Further with the relationship defined by HBI, i. e. $\delta = (\alpha t)^{1/2}[2n(n+1)]^{1/2}$ we have:

$$\frac{\delta}{\sqrt{\alpha t}} = n\sqrt{\pi} = \sqrt{2n(n+1)} \Rightarrow n = \frac{2}{\pi - 2} \approx 1.75 \tag{20}$$

*Example 2.* Prescribed flux boundary condition at $x = 0$

Similarly to the previous problem both temperature profiles at issue are HBIM approximate solution of eq. (3a) with $-\lambda(\partial T/\partial x) = F$ at $x = 0$ and $T(x, 0) = T_\infty$ is [11]:

$$T = T_\infty + \frac{F\delta}{\lambda n}\left(1 - \frac{x}{\delta}\right)^n = T_\infty + \frac{F\delta}{\lambda n}\left(1 - \frac{x}{\sqrt{\alpha t}\sqrt{n(n+1)}}\right)^n \tag{21a, b}$$

with

$$\delta = \sqrt{\alpha t}\sqrt{n(n+1)} \tag{21c}$$

and the exact solution [11, 24]:

$$T_e(x,t) = \frac{2F}{\lambda}\sqrt{\alpha t}\, \text{ierfc}\frac{x}{2\sqrt{\alpha t}} \quad (22)$$

$$= \frac{2F}{\lambda}\sqrt{\alpha t}\left[\frac{1}{\sqrt{\pi}}\exp\left(-\frac{x^2}{4\alpha t}\right) - \frac{x}{2\sqrt{\alpha t}}\left(1-\text{erf}\frac{x}{2\sqrt{\alpha t}}\right)\right]$$

Then the locale entropy generations are as follows:

$$\sigma(T,T_x) = \frac{\lambda}{T_a^2}(\text{grad}\,T_a)^2 = \lambda\frac{\left(\frac{F}{\lambda}\right)^2\left[1-\frac{x}{\delta}\right]^{2n(n-1)}}{\left[T_\infty + \frac{F\delta}{\lambda n}\left(1-\frac{x}{\delta}\right)^n\right]^2} \quad (23a)$$

$$\sigma(T,T_x)_e = \frac{\lambda}{T_e^2}(\text{grad}\,T_e)^2 =$$

$$= \lambda\frac{4\left(\frac{F}{\lambda}\right)^2\alpha t\left[\frac{2x}{\sqrt{\pi}(4\alpha t)} - \frac{1}{2\sqrt{\alpha t}}\text{erf}\frac{x}{2\sqrt{\alpha t}} - \frac{x}{2\sqrt{\pi}(2\alpha t)}\exp\left(-\frac{x^2}{4\alpha t}\right)\right]^2}{\left[\frac{2F}{\lambda}\sqrt{\alpha t}\,\text{ierfc}\frac{x}{2\sqrt{\alpha t}}\right]^2} \quad (23b)$$

At $x = 0$ the condition $\sigma_{x=0}(T,T_x)_a = \sigma_{x=0}(T,T_x)_e$ simply yields:

$$\lambda\frac{\left(\frac{F}{\lambda}\right)^2}{[T_{a(x=0)}]^2} - \lambda\frac{\left(\frac{F}{\lambda}\right)^2}{[T_{e(x=0)}]^2} = 0 \Rightarrow [T_{a(x=0)}]^2 - [T_{e(x=0)}]^2 = 0 \quad (24a)$$

In fact, from (24a) it follows that we have to satisfy the condition:

$$T_{a(x=0)} = T_{e(x=0)} \quad (24b)$$

The condition (24b) is exactly expressed as:

$$T_\infty + \frac{F\delta}{\lambda n} = T_\infty + \frac{2F}{\lambda\sqrt{\pi}}\sqrt{\alpha t} \Rightarrow \frac{\delta}{\sqrt{\alpha t}} = \frac{2n}{\sqrt{\pi}} \quad (25a,b)$$

The result (25b) and that coming from HBIM solution (21c) simply give:

$$\frac{\delta}{\sqrt{\alpha t}} = \frac{2n}{\sqrt{\pi}} = \sqrt{n(n-1)}\sqrt{\frac{n}{n-1}} \Rightarrow \frac{\sqrt{\pi}}{2} = n \Rightarrow \frac{\pi}{4} \approx \pi \approx 3.65 \quad (26)$$

Recall, the condition (24b) was used in [11] as additional physical constraint assuring the definition of the appropriate exponent of the parabolic profile.

**Comments to the test performed with examples 1 and 2**

The condition $\Delta\sigma = \sigma(T,T_x)_a - \sigma(T,T_x)_e = 0, x = 0$ practically means that:

$$\Delta\sigma_{x=0} = \lambda\left[\left(\frac{\partial T_a}{\partial x}\right)^2 - \left(\frac{\partial T_e}{\partial x}\right)^2\right]_{x=0} = 0 \quad (27)$$

In the case of PT problems ($T = T_s$, $x = 0$) we have $T_{a(x=0)} = T_{e(x=0)} = T_s$ and under optimal conditions we search for minimization of $\Delta\sigma$. To this end, the expression (27) is equivalent to:

$$\Delta\sigma \propto \left[\left(\frac{\partial T_a}{\partial x}\right)^2 - \left(\frac{\partial T_e}{\partial x}\right)^2\right]_{x=0} = \left(\frac{\partial T_a}{\partial x} - \frac{\partial T_e}{\partial x}\right)\left(\frac{\partial T_a}{\partial x} + \frac{\partial T_a}{\partial x}\right) = 0 \quad (28)$$

From (28) the only rational condition is:

$$\left.\frac{\partial T_a}{\partial x}\right|_{x=0} = \left.\frac{\partial T_e}{\partial x}\right|_{x=0} \quad (29)$$

which is equivalent to $q_a(x=0) = q_e(x=0)$ and provides the relationship $\delta/(\alpha t)^{1/2} = n\pi^{1/2} = [2n(n+1)]^{1/2}$.

Therefore, *the requirement to minimize the difference in the entropy generation rates* at the point where they exhibit their maxima is physically equivalent to the condition *the heat fluxes at the same point to be equal irrespective of the temperature profiles used in the calculations*. In this context, to avoid some ambiguities, we recall that the volumetric entropy generation based on a volume (see fig. 1c) matching the thermal penetration layer may be expressed through a reference temperature $T_0$ as:

$$S_{gen} \propto \frac{\lambda}{T_0^2}\left(\frac{\partial T}{\partial x}\right)^2 \quad (30)$$

The reference temperature could be either $T_s$ or $T_\infty$. Thus, the only variable of interest representing the optimal approximating profile remain in the ($\partial T/\partial x$)$_a$ expressed through it. This approach leads to (27) and with the condition minimizing the difference in the entropy generation rates (see (16) and (28)) at $x = 0$ provides the optimal value of the profile exponent.

Further, the prescribed flux problem, through the condition $\sigma_{x=0} = 0$ generates the condition $T_{a(x=0)} = T_{e(x=0)} = T_s$ used in [11] and conceived from a physical point of view.

In general, the requirement to minimize the difference in the thermal entropy generations by the approximate and the exact profiles automatically generates additional physical constraints defining the optimal exponent of the parabolic profile at issue. This is an important result, since the condition $\sigma_{x=0} = 0$ inherently contains the condition to minimize the difference in the surface temperatures with prescribed flux and *vice versa*. The main issue is that both additional constraints come automatically through application of a unique physically based condition; to minimize the difference in TEGs at the point where the profiles exhibit their maxima.

**Results outline and ideas thereof**

This point is commonly named "Conclusions" but we will slightly extend its content beyond a brief notation of the principle results of the work and drawing some basic principle and ideas coming from this research note.

The simple idea of EGM through the approximate profile of HBIM was simply tested with two classical examples with known exact solutions. A basic condition derived from this general principle is the requirement the local entropy generation rates calculated through both profiles to be the same at $x = 0$. This condition work very well and confirms results obtained through constraints [11] based on additional physical assumptions.

Further, the requirement $\Delta\sigma = \sigma(T, T_x)_a - \sigma(T, T_x)_e = 0$, $x = 0$ automatically generates the condition which is deficient in the set provided by the HBIM; in case of PF problem and $q_a(x=0) = q_e(x=0)$ with PT problem. With this additional condition the exponent of the parabolic profile can be defined definitely.

Beyond, the results obtained here a principle question rises: Is it possible to apply the method to more complex problems? The task is to create simple HBIM solutions which might replace the huge exact solutions and replace the time-consuming numerical procedures thus providing useful sub-routine functions of complex computational fluid dinamics codes. Certainly, the first results here are promising and a successful step forward is the condition $\sigma(T, T_x) = 0$, $x = 0$ used in this work. The latter simply implies that even though the vast analytical expressions of the exact solutions [24] the approach $\Delta\sigma(T, T_x) = 0$, $x = 0$ needs only the values of $T(0, t)$ and $T_s(0, t)$ which are always simple expressions.

Finally, a more general question rises: Does the approach work with other profiles, especially the so-called "hybrid profiles" [9, 25-27] where a special function with a tuning parameter is used as a term multiplying the basic HBIM profile? To this end, we only release the idea but solutions of specific problems are beyond the present work.

**Acknowledgments**

**Nomenclature**

- $a$ – coefficient in the prescribed temperature profile, [K]
- $b$ – coefficient in the prescribed temperature profile, [K]
- $c$ – coefficient in the prescribed temperature profile, [m$^{-1}$]
- $F(t)$ – surface flux , [Wm$^{-2}$]
- $g$ – volumetric energy source, [Wm$^{-3}$]
- $h_0$ – slab thickness – see eq. (7) , [m]
- $l_0$ – slab width – see eq. (7) , [m]
- $n$ – exponent in the prescribed temperature profile , [–]
- $q_a$ – surface heat flux provided by the approximate temperature profile, [Wm$^{-2}$]
- $q_e$ – surface heat flux provided by the exact temperature profile, [Wm$^{-2}$]
- $S_{gen}$ – entropy generation rate, [WK$^{-1}$]
- $S_{gen}$ – volumetric entropy generation rate, [Wm$^{-3}$K$^{-1}$]
- $\bar{S}_{gen}$ – dimensionless volumetric entropy generation rate (see eq. 7), [–]
- $T$ – temperature, [K]
- $T_e$ – temperature defined by the exact solution, [K]
- $T_\infty$ – temperature of the undisturbed medium, [K]
- $t$ – time, [s]
- $V$ – fluid velocity, [ms$^{-1}$]
- $W_{lost}$ – lost work, [W]
- $x$ – co-ordinate, [m]

*Greek letters*

- $\alpha$ – thermal diffusivity, [m$^2$s$^{-1}$]
- $\lambda$ – thermal conductivity, [Wm$^{-1}$K]
- $\Theta$ – dimensionless temperature $[= (T - T_\infty)/(T_s - T_\infty)]$ [see eqs. (3b) and (7)], [–]
- $\delta$ – thermal layer depth, [m]
- $\sigma(T,T_s)$ – local entropy generation rate (see eq. (10)), [Wm$^{-3}$K$^{-1}$]
- $\sigma_r(T,T_x)$ – global entry generation rate, , [WK$^{-1}$]
- $\Delta\sigma$ – difference in local entropy generation rates (see eq. 16), [Wm$^{-3}$K$^{-1}$]

*Subscripts*

- a – approximate
- e – exact
- p – penetration
- exact – exact solution
- approximate – approximate solution

*Special symbols*

- $\rightarrow$ – it follows that
- – can be expresses as

*Abbreviations*

- EGM – entropy generation minimization
- HBI – heat-balance integral

HBIM – heat-balance integral method  
PF – prescribed flux  
PT – prescribed temperature  
TEG – thermal entropy generation

**References**

[1] Goodman, T. R, The heat-balance Integral and its Application to Problems Involving a Change of Phase, *Transactions of ASME, 80* (1958), 1-2, pp. 335-342
[2] Goodman, T. R., Application of Integral Methods to Transient Nonlinear Heat Transfer, in: Advances in Heat Transfer (Eds. T. F. Irvine, Jr., J. P. Hartnett), Vol. 1, 1964, Academic Press, San Diego, Cal., USA, pp. 51-122
[3] Wood, A. S., Kutluay, S., A Heat Balance Integral Model of the Thermistor, *Int. J. Heat Mass Transfer, 38* (1995), 10, pp. 1831-1840
[4] Moghtaderi, B., *et al.*, An Integral Model for Transient Pyrolysis of Solid Materials, *Fire and Materials, 21* (1997), 1, pp. 7-16
[5] Staggs, J. K. J., Approximate Solutions for the Pyrolysis of Char Forming and Filled Polymers under Thermally Thick Conditions, *Fire and Materials, 24* (2000), 6, pp. 305-308
[6] Ren, H. S., Application of the Heat-Balance Integral to an Inverse Stefan Problem, *Int. J. Thermal Sciences, 46* (2007), 2, pp. 118-127
[7] Mosally, F., Wood, A. S., Al-Fhaid, A., On the Convergence of the Heat Balance Integral Method, *Applied Mathematical Modelling, 29* (2005), 10, pp. 903-912
[8] Theuns, E., *et al.*, Critical Evaluation of an Integral Model for the Pyrolysis of Charring Materials, *Fire Safety Journal, 40* (2005), 2, pp. 121-140
[9] Sahu, S. K., Das, P. K., Bhattacharyya, S., A Comprehensive Analysis of Conduction-Controlled Rewetting by the Heat Balance Integral Method, *Int. J. Heat Mass Transfer, 4*9 (2006), 25-26, pp. 4978-4986
[10] Hristov, J. Y., An Inverse Stefan Problem Relevant to Boilover: Heat Balance Integral Solutions and Analysis, *Thermal Science, 11* (2007), 2, pp. 141-160
[11] Hristov, J. Y., The Heat-Balance Integral Method by a Parabolic Profile with Unspecified Exponent: Analysis and Benchmark Exercises, *Thermal Science, 13* (2009), 2, 27-48 (present issue)
[12] Braga, W., Mantelli, M., A New Approach for the Heat Balance Integral Method Applied to Heat Conduction Problems, *Proceedings*, 38[th] AIAA Thermophysics Conference, Toronto, Ont., Canada, 2005, Paper AIAA-2005- 4686
[13] Langford, D., The Heat Balance Integral Method, *Int. J. Heat Mass Transfer, 16* (1973), 12, pp. 2424-2428
[14] Prigogine, I., Introduction to Thermodynamics of Irreversible Processes, Ch. C. Thomas Publisher, Springfield, Ill., USA, 1955
[15] Bejan, A., Entropy Generation through Heat and Fluid Flow, 1[st] ed., Wiley-Interscience, New York, USA, 1982
[16] Bejan, A., Second Law Analysis in Heat Transfer and Thermal Design, *Advances in Heat Transfer, 15* (1982), 1, pp. 1-58
[17] Bejan, A., Entropy Generation Minimization, 1[st] ed., CRC Press, Boca Raton, Fla., USA, 1996
[18] Zhou, S., Chen, L., Sun, F., Constructal Entropy Generation Minimization for Heat and Mass Transfer in Solid-Gas Reactor Based on Triangular Element, *J. Appl. Phys.D., 40* (2007), 11, pp. 3545-3550
[19] Zhou, S., Chen, L., Sun, F., Optimization of Constructal Volume-Point Conduction with Variable Cross-Section Conducting Path, *Energy Conversion. Management, 48* (2007), 1, pp. 106-111
[20] Yilbas, B. S., Entropy Analysis of Concentric Annuli with Rotating Outer Cylinder, Exergy, *International Journal, 1* (2001), 1, pp. 60-61
[21] Pakdemirli, M., Yilbas, B. S., Entropy Generation for Pipe Flow of Third Grade Fluid with Vogel Model Viscosity, *Int. J. Non-Linear Mechanics, 41* (2006), 3, pp. 432-437
[22] Esfahani, J. A., Roohi, E., Entropy Generation Analysis in Error Estimation of Appropriate Solution Methods Implemented to Conductive Problem, *Proceedings*, 18[th] National and 7[th] ISHMT-ASME Heat and Mass Transfer Conference, IIT Guwahati, India, 2006, Paper N: G 484, 2006
[23] Kolenda, Z., Donizak, J., Hubert, J. On the Minimum Entropy Production in Steady-State Heat Conduction Processes, *Energy, 29* (2004), 12-13 , pp. 2441-2460
[24] Carslaw, H. S., Jaeger, J. C., Conduction of Heat in Solids, 2[nd] ed., Oxford Science Publications, Oxford University Press, Oxford, UK, 1992 (a photocopy edition of 2[nd] ed., 1959)

[25] Venkateshan, S. P., Kothari, N. S., Approximate Solution of One-Dimensional Heat Diffusion Problems Via Hybrid Profiles, *Int. J. Heat and Fluid Flow, 8* (1987), 3, pp. 243-247
[26] Venkateshan, S. P., Solaiappan, O., Approximate Solution of Nonlinear Transient Heat Conduction in One Dimension, *Heat Mass Transfer, 26* (1988), 2, pp. 229-233
[27] Venkateshan, S. P., Rao, V. R., Approximate Solution of Non-Linear Transient Heat Conduction in Cylindrical Geometry, *Heat Mass Transfer, 26* (1991), 1, pp. 97-102

Authors´ affiliation:

*J. Hristov*
Department of Chemical Engineering,
University of Chemical Technology and Metallurgy
1756 Sofia, 8, Kl. Ohridsky Blvd., Bulgaria
E-mail: jordan.hristov@mail.bg